\documentclass[showpacs,prl,floats,aps,superscriptaddress,twocolumn,tightenlines]{revtex4}
\usepackage{amsmath,amssymb,amsfonts}

\usepackage{graphicx}

\begin{document}
\preprint{UTBRG-2001-004, gr-qc/0205043}

\title{Computing the gravitational self-force on a compact object
plunging into a Schwarzschild black hole}

\author{Leor Barack}
\affiliation{Albert-Einstein-Institut, Max-Planck-Institut f{\"u}r Gravitationsphysik,
Am M\"uhlenberg 1, D-14476 Golm, Germany}
\author{Carlos O. Lousto}
\affiliation{Department of Physics and Astronomy,
The University of Texas at Brownsville, Brownsville, Texas 78520}
%\affiliation{Instituto de Astronom\'{\i}a y F\'{\i}sica del Espacio--CONICET,
%Buenos Aires, Argentina}

\date{\today}

\begin{abstract}
We compute the gravitational self-force (or ``radiation reaction''
force) acting on a particle falling radially into a Schwarzschild
black hole.  Our calculation is based on the ``mode-sum'' method, in
which one first calculates the individual $\ell$-multipole
contributions to the self-force (by numerically integrating the
decoupled perturbation equations) and then regularizes the sum over
modes by applying a certain analytic procedure.  We demonstrate the
equivalence of this method with the $\zeta-$function scheme.  The
convergence rate of the mode-sum series is considerably improved here
(thus reducing computational requirements) by employing an
analytic approximation at large $\ell$.

%The convergent rate of the mode-sum series is considerably
%improved here by introducing an analytic approximation at large $\ell$
%notably reducing the computational requirements.
\end{abstract}

\pacs{04.25.Nx, 04.30.Db, 04.70.Bw}
\maketitle

%--------------------------------------------------------------------

%\P {\bf Motivation and the recent availability of SF calculations schemes:}

The space-based gravitational wave detector LISA (Laser Interferometer
Space Antenna), scheduled for launch around 2011 \cite{LISA}, will
open up a window for the low frequency band below 1Hz, allowing access
to a variety of black hole sources. A main target for LISA would be
the outburst of gravitational radiation emitted during the capture of
a compact star by a supermassive black hole---a $10^{5-7}$ solar
masses black hole of the kind now believed to reside in the cores of
many galaxies, including our own \cite{Nature}. By LISA's launch time,
a sufficient theoretical understanding of the orbital evolution of
such systems, including radiation reaction effects, must be at hand,
to allow design of accurate templates necessary for detection and
interpretation of the gravitational waveforms.  Due to the extreme
mass-ratio typical to the binary systems of interest, the entire
problem can be conveniently treated in the context of perturbation
theory, being a relatively mature branch of gravitational physics: the
compact stellar object is thus modeled by a point-like particle, and
its field treated as a perturbation over the fixed Kerr geometry of
the large black hole. To leading order in the mass ratio, such a
particle then traces a geodesic of the background spacetime, and one
asks about radiation-reaction induced corrections to this geodesic.

The ``traditional'' approach to this problem relies on energy-momentum
balance considerations \cite{H00}: By calculating the fluxes of energy
and angular momentum to infinity and across the event horizon, one
attempts to infer the temporal rate of change of the particle's
``constants'' of motion. This technique is applicable only in
adiabatic scenarios, in which the time scale for radiation reaction
effect is much larger than the dynamic time-scale of the system; it is
not clear yet whether this approximation is valid for the entire range
of relevant LISA parameters
\cite{H00}. Moreover, this approach seems insufficient for
tackling generic orbits in Kerr spacetime (i.e., ones both eccentric
and inclined) even under the adiabatic approximation.
%: tracking the
%evolution of such orbits requires knowledge of the rate of change
%of a third ``constant'' of motion---the Carter constant---which, alas, is
%not imprinted as a ``flux'' at infinity (this constant is related in a
%nonlinear way to the angular momentum's components) \cite{H00}.
%Hence, accounting for radiation reaction effect in systems of actual
%astrophysical interest seems to oblige one to turn to the notion of
%{\it local self force} (SF)
This has led many researches, particularly over the last five years,
to turn to the useful notion of the {\it local self-force} (SF).

Consider a point-like particle of mass $\mu$ moving freely around
a black hole with mass $M\gg\mu$; and treat the particle's gravitational
field, $h_{\alpha\beta}\propto O(\mu)$, as a linear perturbation over
the background metric $g_{\alpha\beta}$. In the ``SF picture'', this
particle's equation of motion is written as
\begin{equation} \label{EOM}
\mu u^{\alpha}_{\ ;\beta}u^{\beta}=F^{\alpha}_{\rm self},
\end{equation}
where $u^{\alpha}$ is the particle's four-velocity, a semicolon denotes
covariant differentiation with respect to  $g_{\alpha\beta}$, and
$F^{\alpha}_{\rm self}\propto O(\mu^2)$ describes the leading-order
SF effect. Since the perturbation $h_{\alpha\beta}$ obviously diverges
at the particle's location,
the problem of obtaining $F^{\alpha}_{\rm self}$
involves the introduction of a reliable regularization scheme.
A well established formal prescription for constructing
$F^{\alpha}_{\rm self}$, relying on a physically consistent regularization
method, became available recently with the work of Mino, Sasaki and Tanaka
(MST) \cite{MST97}.
The same formal prescription was introduced independently by Quinn and
Wald (QW) \cite{QW97}, based on an axiomatic approach.
To allow a practical implementation of the
MST/QW formal prescription in actual calculations,
a ``mode-sum'' scheme was later devised in
Refs.~\cite{MSRS-scalar,B01,BMNOS02},
based on the MST/QW result. An alternative regularization
approach, based on the $\zeta-$function technique, was introduced in
Ref.~\cite{L00}.

The main objective of this paper is to report on a first actual
calculation of the gravitational SF, based on the mode-sum
prescription. Focusing, as a test-case, on radial trajectories
in a Schwarzschild background, we
demonstrate the applicability of this approach, and push forward some
analytic and numerical techniques which may later be applied to more
general orbits. Among the new results presented here: 
(i) Two different derivations of the ``regularization parameters'',
independent of each other and of the derivation
of \cite{BMNOS02};
(ii) Consistency of the MST/QW regularization with the $\zeta$-function method;
(iii) A first explicit example of the gauge-invariance feature of the
regularization parameters predicted in Ref.~\cite{BO01}; 
(iv) An improved numerical method for integrating the decoupled field
equations to fourth order accuracy;
(v) An analytic approximation developed for improving the convergence
rate of the mode-sum series (see below).
%, and even allowing to approximate the SF analytically. 
Full details of our analysis %on its various parts,
shall be provided elsewhere \cite{BL02}.

Throughout this paper we use ``geometrized'' units $G=c=1$, metric
signature ${-}{+}{+}{+}$, and the standard Schwarzschild coordinates
$t,r,\theta,\varphi$. We consider a particle falling radially
into a Schwarzschild black hole with mass $M$, starting at rest at
$r=r_0$, and let $r_p$ denote the value of $r$ at the SF evaluation
point. In the lack of SF, the particle traces a geodesic
characterized by the (conserved) specific energy $E\equiv (1-2M/r_0)^{1/2}$.
By virtue of the symmetry of the above setup, we obviously have
$F_{\rm self}^\theta=F_{\rm self}^\varphi=0;$ we hereafter thus focus on the
$r,t$ components of the SF.
%--------------------------------------------------------------------

Let us start by briefly reviewing the mode-sum method for
constructing the gravitational SF:
First, one has to calculate the multipole modes of the
metric perturbation $h_{\alpha\beta}$ in the harmonic gauge, denoted here by
$h_{\alpha\beta}^\ell$ (this refers to the quantity obtained by summing
over all azimuthal numbers $m$ and over all ten tensor harmonics for
a given multipole number $\ell$). This calculation is done through a
numerical integration of the decoupled linearized Einstein equations.
%(e.g., in the Regge-Wheeler gauge \cite{RW57,Z70}, as in our current analysis)
Then, one constructs the $\ell$-mode contribution to the
``full'' force, denoted here by $F^{\alpha\ell}$, through a certain
operation involving 1st-order derivatives of $h_{\alpha\beta}^\ell$
[see Eq.\ (15) of Ref.\ \cite{BMNOS02}].
In the radial motion case, this operation reduces to
\begin{equation} \label{Fl}
F^{\alpha\ell}_{\pm}=\mu\, k^{\alpha\beta\gamma\delta}
\bar{h}^\ell_{\beta\gamma;\delta}
\end{equation}
(evaluated at the particle's location), where
%$\bar{h}^\ell_{\beta\gamma}$ are the ``trace-reversed'' modes,
$\bar h^\ell_{\alpha\beta}\equiv
h^\ell_{\alpha\beta}-\frac{1}{2}g_{\alpha\beta}
g^{\mu\nu}h^\ell_{\mu\nu}$, $k^{\alpha\beta\gamma\delta}\equiv
u^{\beta}u^{\gamma}g^{\alpha\delta}/2+g^{\beta\gamma}g^{\alpha\delta}/4
+u^{\alpha}g^{\beta\gamma}u^{\delta}/4-g^{\alpha\beta}u^{\gamma}u^{\delta}
-u^{\alpha}u^{\beta}u^{\gamma}u^{\delta}/2$, and the $\pm$ sign
%assigned to the modes $F_{\alpha}^{l}$
corresponds to taking the derivative %in Eq.\ (\ref{Fl})
from $r\to r_p^{\pm}$, respectively.
%, where $r=r_p$ is the radial location of the particle.
(Note that these force-modes satisfy the normalization condition
$u_\alpha F^{\alpha\ell}_\pm=0$.)
%We comment here that, 
While the
perturbation itself diverges at the particle's location, the
individual modes $h^\ell_{\alpha\beta}$ are continuous
everywhere~\cite{L00}---an important benefit of the mode-sum approach
(this holds in the harmonic gauge or in any other gauge related to it
by a regular gauge transformation). Typically, however, the
derivatives of $h^\ell_{\alpha\beta}$ are found to have a finite
discontinuity through the particle's location, yielding two different
finite values $F^{\alpha\ell}_{\pm}$.  According to the mode-sum
method, the gravitational SF is then constructed
through~\cite{B01}
\begin{equation}\label{MSRS}
F^{\alpha}_{\rm self}=\sum_{\ell=0}^{\infty}\left[F^{\alpha\ell}_{\pm}
-A_{\pm}^{\alpha}L-B^{\alpha}-C^{\alpha}/L\right] -D^{\alpha},
\end{equation}
where $L\equiv \ell+1/2$ and the ($\ell$-independent) quantities
$A^{\alpha}$, $B^{\alpha}$, $C^{\alpha}$, and $D^{\alpha}$ are the
so-called ``regularization parameters'', whose values depend on the
orbit under consideration. Roughly speaking, the expression
$A_{\pm}^{\alpha}L+B^{\alpha}+C^{\alpha}/L$ reflects the asymptotic
form of $F^{\alpha\ell}_{\pm}$ at large $\ell$ [ensuring convergence
of the sum in Eq.\ (\ref{MSRS})], while the parameter $D^{\alpha}$ is
a certain residual quantity that arises in the summation over $\ell$.
(See \cite{B01,BMNOS02} for an exact definition of these parameters.)

%--------------------------------------------------------------------

Incorporating a systematic perturbation expansion of the 
$\ell-$mode Green's
function associated with the perturbation equations in the harmonic
gauge---an implementation of the technique developed in
\cite{B01}---we have obtained for radial trajectories~\cite{BL02},
%~~~~~~~~~~~~~~~~~~~~~~~~~~~~~~~~~~~~~~~~~~~~~~~~~~~~~~~~~~~~~~~~
\begin{subequations} \label{RP}
\begin{equation}\label{A}
A_{\pm}^r=\mp \frac{\mu^2}{r_p^2}\,E,
\quad\quad
A_{\pm}^t=\mp \frac{\mu^2}{r_p^2}\,\frac{\dot{r}_p}{f},
%\quad
%A_{\pm}^\theta=A_{\pm}^\varphi=0,
\end{equation}
\begin{equation}\label{B}
B^r=-\frac{\mu^2}{2r_p^2}\,E^2,
\quad\quad
B^t=-\frac{\mu^2}{2r_p^2}\,\frac{E\dot{r}_p}f,
%\quad
%B^\theta=B^\varphi=0,
\end{equation}
\begin{equation}\label{CD}
C^{\alpha}=D^{\alpha}=0,
\end{equation}
\end{subequations}
%~~~~~~~~~~~~~~~~~~~~~~~~~~~~~~~~~~~~~~~~~~~~~~~~~~~~~~~~~~~~~~~~
where $f\equiv 1-2M/r_p$ and $\dot{r_p}=-(E^2-f)^{1/2}$. 
%These values apply to the modes $F^{\alpha\ell}_{\pm}$
%calculated in the harmonic gauge [i.e. the ones derived, via Eq.~(\ref{Fl}),
%from the harmonic gauge perturbation modes.]
These values %(\ref{RP})
agree with those derived (for generic orbits)
in Ref.~\cite{BMNOS02} using a different method.  Note that whereas
the values of $A_{\pm}^{\alpha}$, $B^{\alpha}$, and $C^{\alpha}$ can
be verified numerically by examining the behavior of the modes
$F^{\alpha\ell}_\pm$ at large $\ell$ (see below), the value of
$D^{\alpha}$ cannot be so verified; hence the
importance of our independent derivation of $D^{\alpha}$.

The above prescription requires one to tackle the perturbation equations
in the harmonic gauge. These equations are separable with respect to $\ell,m$
\cite{B01}, but it is not clear how, or whether at all, one could avoid,
% in this gauge,
the coupling occurring between different elements of the
tensor-harmonic basis.  A more practical derivation of
$h^\ell_{\alpha\beta}$ %the perturbation modes 
(and, consequently, of $F^{\alpha\ \ell}_{\pm}$) is possible in the
{\em Regge-Wheeler} (RW) gauge~\cite{RW57,Z70}: Here, a well
developed formalism~\cite{M74} allows one to derive all
$h^{\ell}_{\alpha\beta}$ components from two scalar generating
functions, by mere differentiation \cite{L00,BL02}. These two waveforms
%functions (yielding, correspondingly, the ``even'' and ``odd'' parity
%parts of the perturbation \cite{RW57,Z70}) each satisfies a
satisfy a scalar-like wave equation which is
easily accessible to numerical treatment.
Now, it has been shown \cite{BO01} that the mode-sum formula
(\ref{MSRS}) is valid, {\em with the same parameter values}, for any
gauge related to the harmonic gauge through a regular gauge
transformation. The RW gauge belongs to this regular family of gauges
so long as radial trajectories are considered
\cite{BO01}---as in our current work. This shall allow us to work here
entirely within the convenient RW gauge.

Using a variant of the Green's function expansion technique mentioned
above---this time applied to the perturbation equations in the RW
gauge---we have been able to directly obtain the values of $A_{\alpha}$,
$B_{\alpha}$, and $C_{\alpha}$ associated with the RW-gauge modes
$F_{\alpha}^{\ell\pm}$.  (The details of this derivation shall be given
in \cite{BL02}.)  The RW-gauge parameters thus obtained were found to
have, in the head-on case considered here, {\em precisely the same
values} as in the harmonic gauge. This serves as a
first explicit demonstration of the regularization parameters'
gauge-invariance property predicted in \cite{BO01}.

%--------------------------------------------------------------------

%It should be noted that the mode-sum prescription (\ref{MSRS}),
%stemming from the standard MST/QW regularization scheme, completely
%conforms, in the head-on case, with the generalized Riemann's
%$\zeta-$function regularization method developed in Ref.~\cite{L00}.
%In the latter the SF can be brought to the form (\ref{MSRS})
%as well, with the
%parameter $D^{\alpha}\propto\zeta(0,1/2)$ shown to vanish (for any orbit)
%using $\zeta-$function regularization arguments~\cite{L01}.
%--------------------------------------------------------------------

It should be noted that the mode-sum prescription (\ref{MSRS}), stemming
from the standard MST/QW regularization scheme, completely conforms
with the $\zeta$-function regularization approach introduced
in Ref.~\cite{L00}: In the latter too, the SF is brought to the
form (\ref{MSRS}), with the parameter $D_{\alpha}$ shown \cite{L00,L01}
to be $\propto\zeta(0,1/2)=0$ (where $\zeta$ is Riemann's generalized 
zeta function).
%Moreover it was shown numerically in Ref.~\cite{L00} and analytically in
%Ref.~\cite{L01} the the average values of $A_\alpha$ and $C_\alpha$
%must vanish. (If $A$ do not vanish, $D$ is not vanishing, if $C$ does
%not vanish, then we cannot regularize!)

Incorporating the parameter values (\ref{RP}) in Eq.~(\ref{MSRS})
(noticing that $A^{\alpha}_{+}=-A^{\alpha}_{-}$), we next write the
mode-sum formula in the compact form
\begin{equation}\label{MSRS2}
F^{\alpha}_{\rm self}
=\sum_{\ell=0}^{\infty}\left(\bar F^{\alpha\ell}-B^{\alpha}\right)
\equiv \sum_{\ell=0}^{\infty} F_{\rm reg}^{\alpha\ell},
\end{equation}
where $\bar F^{\alpha\ell}\equiv (F^{\alpha\ell}_{+}+F^{\alpha\ell}_{-})/2$,
and $F_{\rm reg}^{\alpha\ell}[\propto O(\ell^{-2})]$ are the ``regularized''
modes. Note that $B^{\alpha}(r_p)$ describes the asymptotic form of $\bar
F^{\alpha\ell}$ at $\ell\to\infty$.
%--------------------------------------------------------------------

We now turn to the actual implementation of the prescription
(\ref{MSRS2}), beginning with the numerical calculation of
$\bar{F}^{\alpha\ell}$.  As already mentioned, our calculation was
carried out within the RW gauge. All tensorial components of the
$\ell$-mode metric perturbations in the RW gauge (in fact, only the
even-parity part of $h_{\alpha\beta}$ plays role in our head-on case)
are conveniently constructed from a single scalar generating
function---Moncrief's gauge-invariant waveform $\psi^\ell$
\cite{M74}. This construction, prescribed in \cite{L00}, involves twice
differentiating $\psi^\ell$. Then, the desired modes $F^{\alpha\ell}$
are obtained using Eq.\ (\ref{Fl}).
%(which involves one further differentiation). 
Thus, the numerical task reduces to integrating the
(inhomogeneous) hyperbolic wave equation satisfied by $\psi^\ell$
\cite{LP}, with the appropriate source term associated with the
point-like particle, and with a proper choice of initial data. This
numerical problem has been formalized and worked out previously in
Ref.~\cite{LP}.
%by Lousto and Price \cite{LP}.
We have developed an improved version of the above numerical scheme,
%described in \cite{LP},
which ensures fourth-order numerical convergence.
(This is essential for our purpose, as the construction of $F^{\alpha\ell}$
involves {\em three} derivatives of the numerical-integration variable
$\psi^\ell$.)

\begin{figure}[thb]
%\input{epsf}
%\centerline{\epsfysize 6cm \epsfbox{figure1.eps}}
\begin{center}
\includegraphics[width=3.4in]{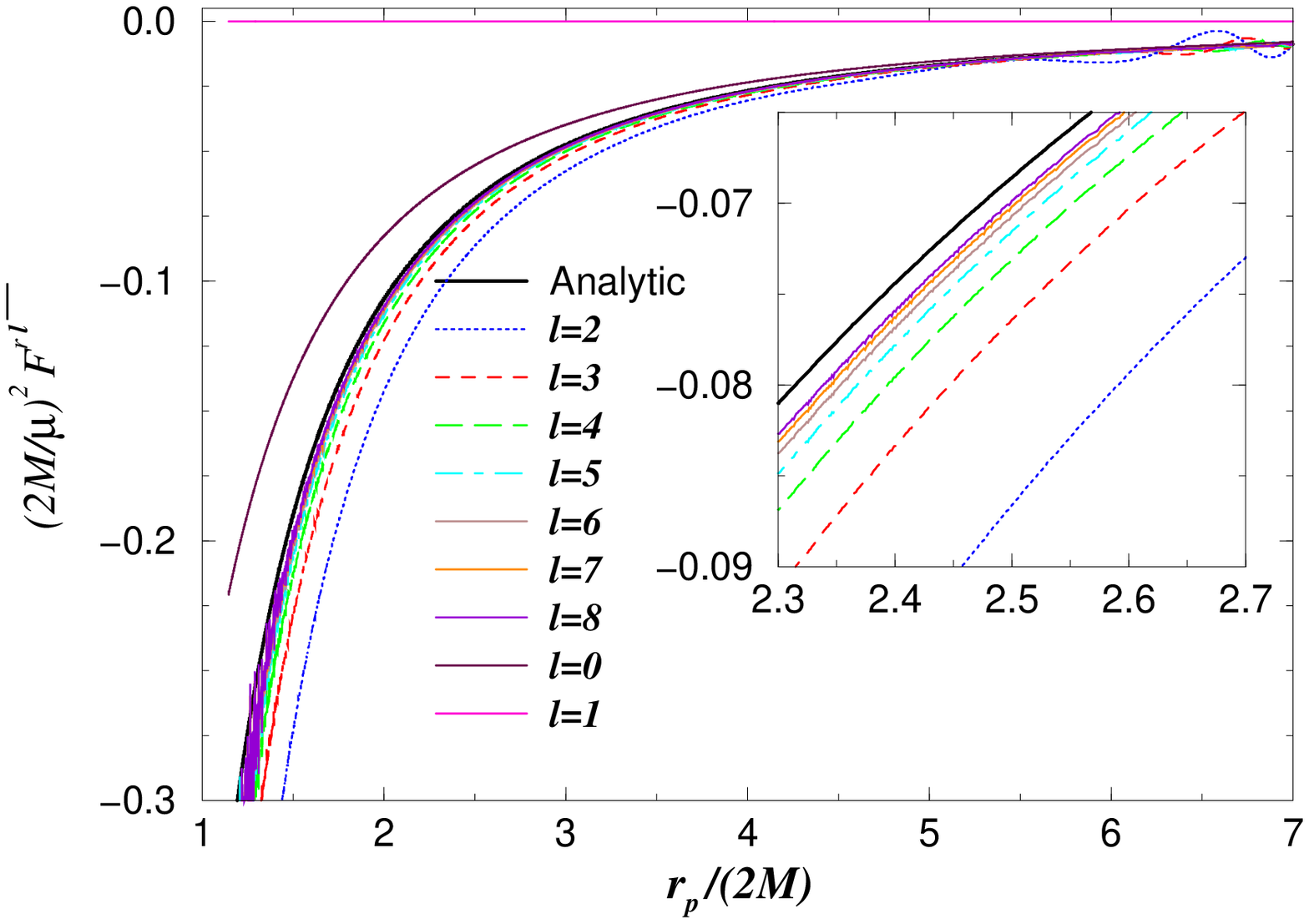}\\
\includegraphics[width=3.4in]{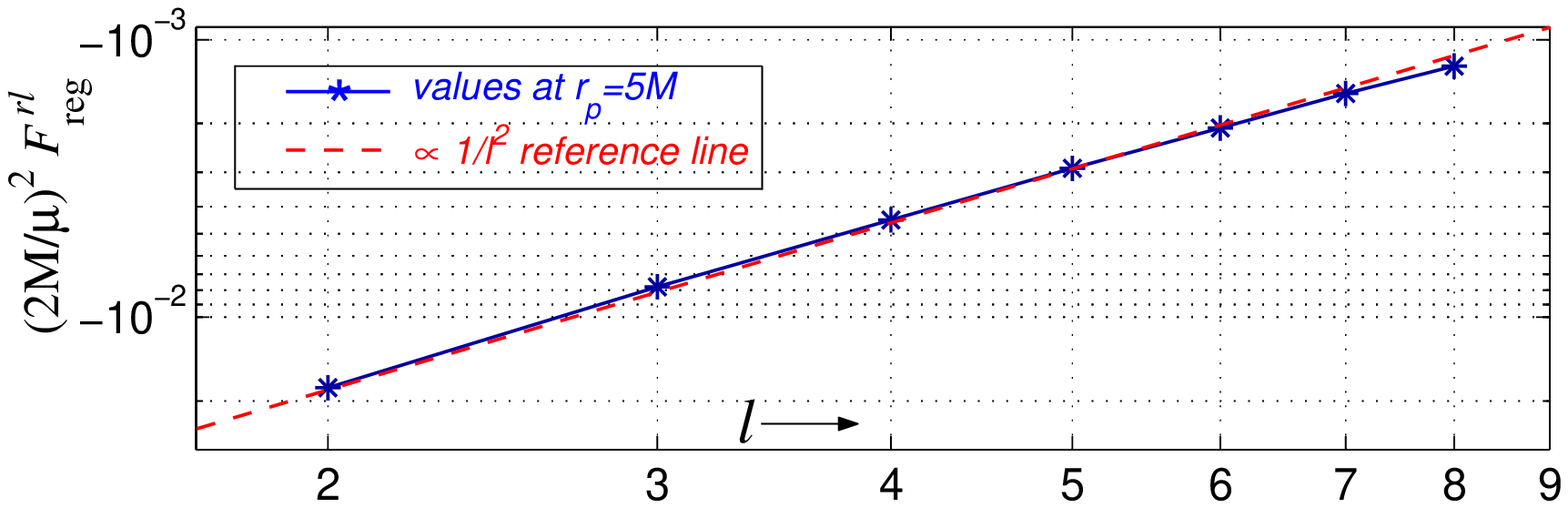}
\end{center}
\caption{%\protect\footnotesize
Upper figure: The ``full'' modes $\bar F^{r\ell}(r_p)$ at
$\ell=0,\ldots,8$, for a particle released from rest at $r_0=14M$.
Note how a limiting curve [given analytically by $B^r(r)$] is approached
at large $\ell$.
%The inset shows how a limiting curve [given by $B_r(r)$] is approached
%at large $\ell$ with a $\propto\ell^{-2}$ behavior. This is
%demonstrated in the lower part of the figure (for a fiducial
%$r_p=5M$).
The wavy feature near $r_p=14M$ is due to the radiation content of
the initial data (chosen here as conformally flat). This ``spureous''
feature damps
down by the time the particle reaches $r_p\sim10M$, exposing the
inherent self force effect.  The bottom figure demonstrates the
anticipated $\propto\ell^{-2}$ convergence of the difference $F_{\rm
reg}^{r\ell}\equiv\bar F^{r\ell}-B^r$.}
%(See Eq.~(\ref{NOr})). }
%component of the self force independent of the initial data.}
\label{figure1}
\end{figure}

Typical results from applying the above numerical prescription are
presented in Fig.\ \ref{figure1}, showing the first few (averaged)
modes $\bar F^{r\ell}$ as a function of $r_p$ for the sample value
$r_0=14M$ ($E\cong 0.926$), and demonstrating the anticipated $\propto
\ell^{-2}$ behavior of %the ``regularized'' modes 
$F_{\rm reg}^{r\ell}$.
The above construction of $\bar{F}^{\alpha\ell}$ %through Moncrief's formalism
is only applicable to the ``radiative'' modes $\ell\geq 2$. 
The modes $l=0,1$, although merely reflecting a residual gauge freedom
\cite{RW57}, must also be taken into account in the mode sum (\ref{MSRS2}).
As shown in \cite{Z70}, the $l=1$ even-parity perturbation is completely
removable by a gauge transformation (interpreted as a translation to the
center of mass system), and we may take $\bar F^{l=1}_{\alpha}=0$.
The $l=0$ perturbation mode 
(interpreted as a variation in the mass $M$)
is constructed analytically in \cite{BL02}---the resultant contribution
$\bar F^{l=0}_{\alpha}$ is also plotted in Fig.\ \ref{figure1}. \cite{l=0}

%~~~~~~~~~~~~~~~~~~~~~~~~~~~~~ FIGURE 1 ~~~~~~~~~~~~~~~~~~~~~~~~~~~~~
%~~~~~~~~~~~~~~~~~~~~~~~~~~~~~~~~~~~~~~~~~~~~~~~~~~~~~~~~~~~~~~~~~~~~
%--------------------------------------------------------------------

\begin{figure}[thb]
\includegraphics[width=3.4in]{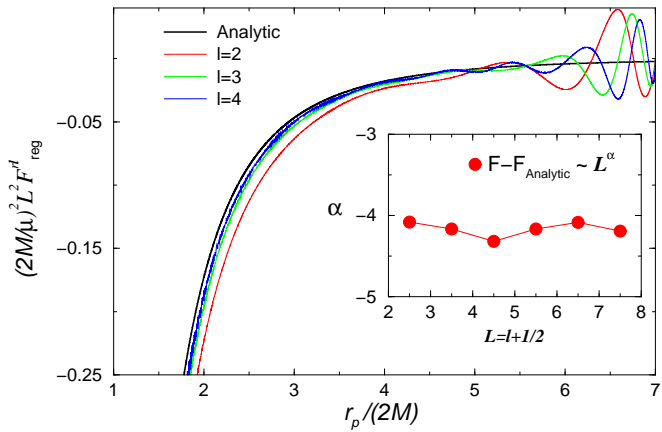}
\caption{
Analytic approximation {\it vs.} numerical results: The plot shows the
numerically calculated (regularized) modes $F_{\rm reg}^{r\ell}$ along
with their large-$\ell$ analytic approximation, $F^{r\ell}_{\rm
analytic}$, as given by the $O(L^{-2})$ term in Eq.\ (\ref{NOr}). Shown
are the modes $\ell=0,1,2,3,4$ for $r_0=14M$ [times $L^2$ for the sake of
comparison].  The inset shows the reminder $F^{r\ell}_{\rm
reg}-F^{r\ell}_{\rm analytic}$ at $r_p=6M$, demonstrating its anticipated
$\propto\ell^{-4}$ behavior.  The wavy feature at the onset of the plunge
is associated with the ``spurious'' radiation content of the initial data;
the inherent SF is exposed only after these waves are dissipated away.}
\label{figure2}
\end{figure}

Since $F_{\rm reg}^{\alpha\ell}\propto O(\ell^{-2})$, the mode sum in
(\ref{MSRS2}) admits the slow convergence rate $\propto 1/\ell$. This
means that achieving even a modest accuracy requires one to sum over
many modes, which is numerically very demanding (numerical integration
of the decoupled field equations becomes increasingly difficult at
growing $\ell$).  To improve the convergence rate of the mode-sum series,
we have obtained \cite{BL02} an analytic approximation for $F_{\rm
reg}^{\alpha\ell}$ at large $\ell$: By extending the local analysis of
the $\ell-$mode Green's function %Moncrief's waveform
one order beyond the calculation of the three
parameters $A^{\alpha}$, $B^{\alpha}$, and $C^{\alpha}$, we have
obtained an analytic expression for
%been able to calculate analytically the ``next order'' in the mode-sum
%expansion, i.e.,
 the $O(L^{-2})$ term of $F^{\alpha\ell}$. This
significantly improves the mode-sum convergence, especially by virtue
of the fact that the next, $O(L^{-3})$ term in the mode-sum is
expected to vanish (this can be shown for any positive odd power of
$1/L$ in the mode sum using straightforward parity arguments
\cite{L01}, and is further supported by our numerical
results---see Fig.\ \ref{figure2}).
We have found \cite{BL02}
\begin{subequations}\label{NO}
\begin{equation} \label{NOr}
F^{r\ell}_{\rm reg}\!=\!
-\frac{15}{16}\mu^2\frac{E^2}{r_p^2}\left(E^2+\frac{4M}{r_p}-1\right)L^{-2}
%\frac{15}{8}\mu^2 (E^2/f)r_p^{-1/2}\frac{d}{d\tau}(\dot{r}_pr_p^{-1/2})L^{-2}
\!+O(L^{-4}),
\end{equation}
\begin{equation} \label{NOt}
F^\ell_{t\ \rm reg}=
-\frac{15}{16}\mu^2 E \frac{d}{d\tau}\left(\frac{\dot{r}_p^2}{r_p}\right)L^{-2}
+O(L^{-4}),
\end{equation}
\end{subequations}
where $\tau$ is the particle's proper time
(the two expressions are negative definite for all $r<r_0$).
These expressions are in perfect agreement with the numerical results,
as demonstrated in Fig.\ \ref{figure2}.
% In particular the difference
%between the analytic approximation and the numerical computation of
%the reaction force scales like  $O(L^{-4})$ as shown in the inset of
%Fig.\ \ref{figure2}. The oscillations near $r=14M$ are due to the
%spurious radiation content of the initial data, and quickly dumps down.
The contribution of the $O(L^{-2})$ expansion term to the overall SF
is now easily obtained analytically, using
$\sum_{l=2}^{\infty}L^{-2}
%=\zeta(2,1/2)-40/9
\cong 0.49$.  The
remainder of the mode sum now converges as $\propto\ell^{-3}$.  By
calculating numerically only the first 10 modes (say), one now obtains the
SF to within a mere relative error of $\sim 10^{-3}$ (compare this with a $\sim
10^{-1}$ error when not using the analytic approximation).
%~~~~~~~~~~~~~~~~~~~~~~~~~~~~~ FIGURE 2 ~~~~~~~~~~~~~~~~~~~~~~~~~~~~~
%~~~~~~~~~~~~~~~~~~~~~~~~~~~~~~~~~~~~~~~~~~~~~~~~~~~~~~~~~~~~~~~~~~~~
%This involves a certain assumption concerning the regularity
%of the tail term, but can be verified numerically so there's no
%problem with that.
%(We comment that Accomplishing this task using the method of
%\cite{BMNOS02} appears difficult... requires next order in the
%Hadamard expansion.)

%Figure \ref{figure3} shows the overall self force 
%(both its $r$ and $t$ components),
%resulting from summing over the individual mode contributions
% (including those of $l=0,1$),
% and incorporating the above analytic approximation for
%improving the convergent rate.% for $\ell\leq9$.

\begin{figure}[thb]
\includegraphics[width=3.4in]{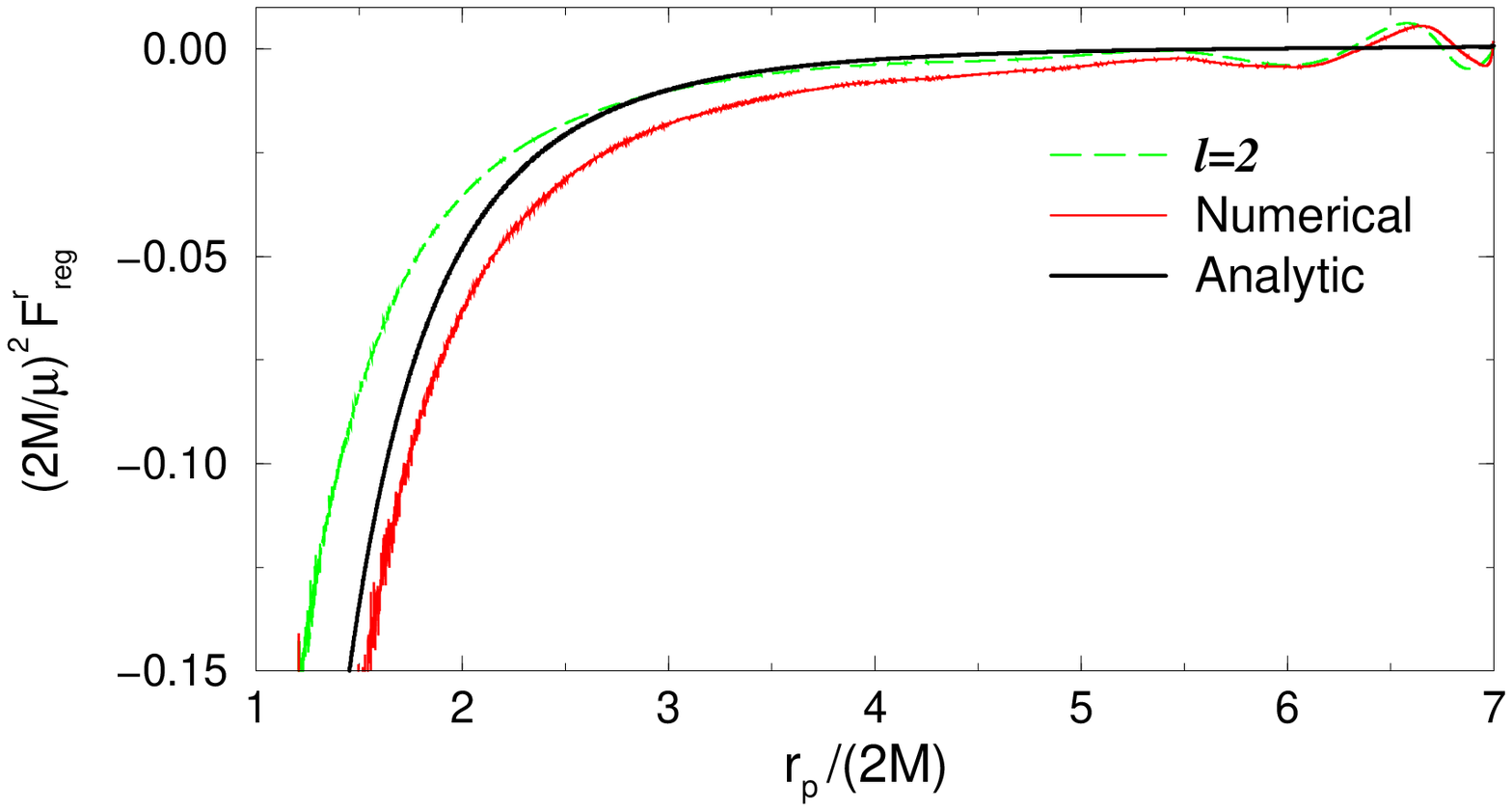}\\
\includegraphics[width=3.4in]{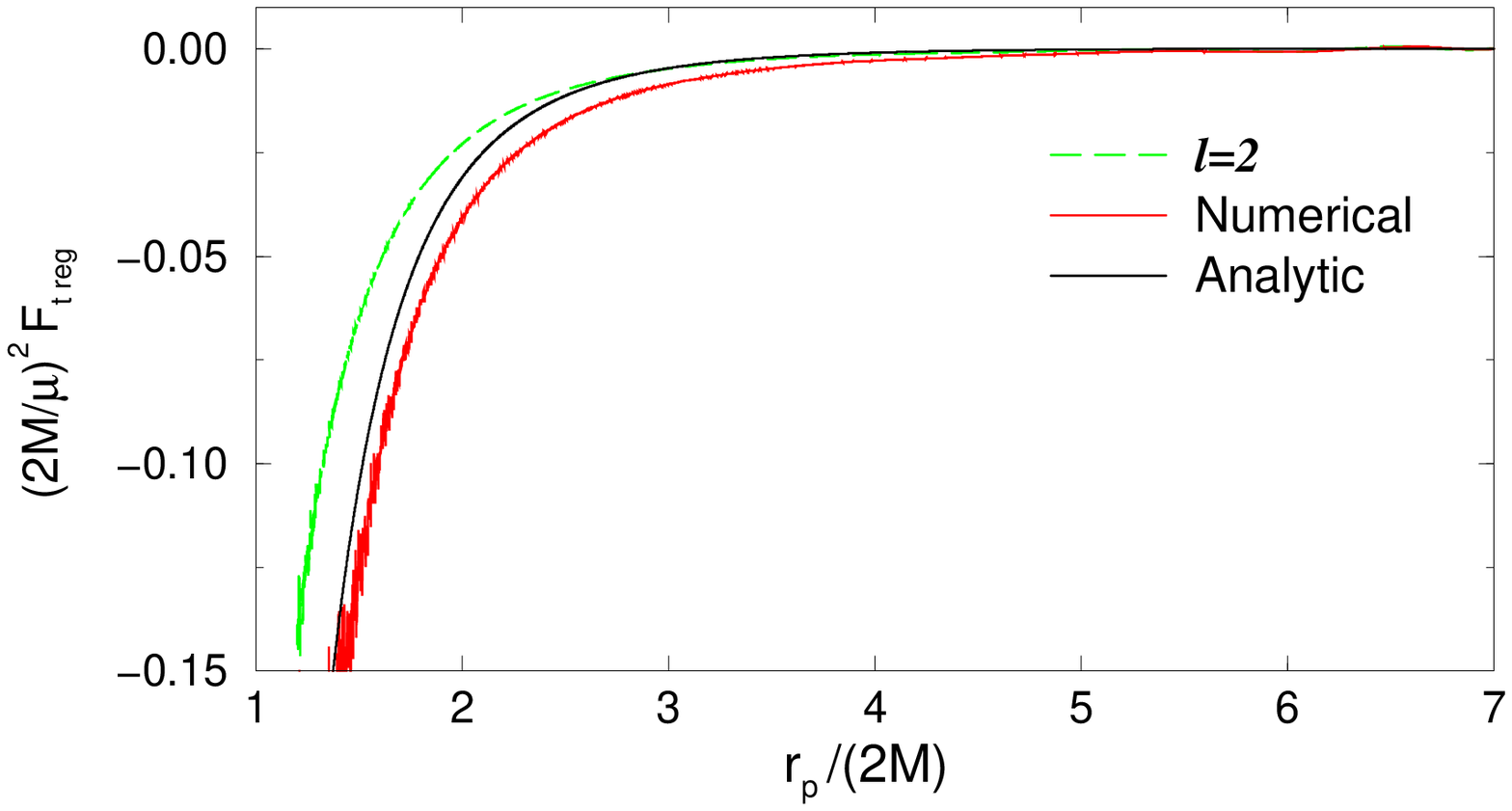}
\caption{
%The upper and bottom
The upper and bottom
panels show, respectively, the $r$ and $t$ components
of the overall SF on a particle starting at rest at $r_0=14M$. The plots
labeled as ``numerical'' 
are produced by summing up the numerically calculated (regularized) modes
up to $\ell=8$, and then incorporating our analytic approximation at $\ell>8$
(these higher modes contribute up to $\sim20\%$ of the total force).
Also given, for comparison, is a curve based entirely on the analytic
approximation (\ref{NO}) (summed over $\ell=2,\ldots,\infty$
plus the exact solutions for $l=0,1$); and a curve
showing the mere $\ell=2$ contribution. The latter serves to illustrate the
importance of higher $\ell$ contributions. All curves reach a finite value
at the horizon.
%The overall ``self force'' on a particle starting at
%rest at $r_0=14M$ ($E=0.925$).
%These plots are produced by summing up the numerically calculated
%(regularized) modes $l=2,\ldots,8$, then continue the sum with the
%analytic approximation. For comparison is also shown the purely
%analytic approximation summing over $l=2,\ldots,\infty$, and solely
%the $\ell=2$ mode to display the importance of the contribution
%from higher $\ell-$modes.
}
\label{figure3}
\end{figure}

Figure \ref{figure3} shows both $r$ and $t$ components of the overall SF 
resulting from summing up all individual mode contributions. The modes
$\ell=2,\ldots,8$ were obtained numerically, while for $\ell> 8$ we used
the analytic approximation (\ref{NO}) (for $\ell=0,1$ we used the exact
solutions mentioned above).
%~~~~~~~~~~~~~~~~~~~~~~~~~~~~~ FIGURE 3 ~~~~~~~~~~~~~~~~~~~~~~~~~~~~~
%~~~~~~~~~~~~~~~~~~~~~~~~~~~~~~~~~~~~~~~~~~~~~~~~~~~~~~~~~~~~~~~~~~~~
The radial component of the SF is found to point {\em inward} (i.e.,
toward the black hole) throughout the entire plunge. This seems to be
a universal feature which does not depend on the starting point $r_0$.
Consequently, the work done by the SF on the particle is positive,
resulting in that the energy parameter $E$ {\it increases} throughout
the plunge.  The instantaneous rate of change of $E$ is given
by \cite{O95} $\mu(dE/d\tau)=-F_t\ (\geq0)$, and the total
change of $E$ % to be denoted $\mu\Delta E$,
is obtained by integrating this
expression along the worldline from $\tau(r_0)$ to $\tau(r=2M)$.
It is important to stress, however, that this result will be attached to
our specific choice of gauge (as opposed to the energy flux at infinity,
which is gauge invariant) \cite{QW99}.

In summary, the mode-sum approach for calculating the gravitational
SF was successfully applied here in the test case of radial motion
in Schwarzschild spacetime. We have also demonstrated the
feasibility of applying an analytic approximation for improving the
mode-sum convergence and even providing a rough estimate to the SF.
This marks a significant milestone in our (still long) way toward
being able to compute the orbital evolution of generic orbits in
Kerr spacetime. The next step along this way, already being considered, is the
implementation of the mode-sum prescription to more general orbits in
Schwarzschild background. This will provide a first opportunity for
validating the SF approach against available calculations
based on the standard energy-momentum balance approach.
%\cite{H00}.

%\begin{acknowledgments}
L.B. wishes to thank Amos Ori for illuminating discussions.
C.O.L. thanks AEI, where part of this research took place, for hospitality. 
L.B.\ was supported by a Marie Curie Fellowship of the European Community
program IHP-MCIF-99-1 under contract number HPMF-CT-2000-00851.
We finally thank all participants of the Capra 5 meeting (PSU) for helpful
comments.
%\end{acknowledgments}

%%%%%%%%%%%%%%%%%%%%%%%%%%%%%%%%%%%%%%%%%%%%%%%%%%%%%%%%%%%%%%%%%%%%%%%%%
%\begin{references}


\begin{thebibliography}{99}
\bibitem{l=0}
The ``standard'' $l=0$ solution constructed by Zerilli~\cite{Z70} 
is pathological at the event horizon.
Here we adopt a different, physically consistent, monopole solution
having $K^\ell=0, H^\ell_0=H^\ell_1=H^\ell_2$, in the RW notation~\cite{RW57}.
We shall discuss this issue in more detail in \cite{BL02}.
\end{thebibliography}
\end{document}